\documentclass[aps,prb,twocolumn,superscriptaddress,amsmath,amssymb,showpacs]{revtex4-1}
\usepackage{amssymb}
\usepackage{amsmath}
\usepackage{amsfonts}
\usepackage{ dsfont }
\usepackage{graphicx}
\usepackage{bm}
\usepackage{subfigure}
\usepackage{epsfig}
\usepackage{color}
\usepackage{bbding}
\usepackage{array,multirow}
\usepackage{soul}
\usepackage[ansinew]{inputenc}

\usepackage{float}
\usepackage{color}
\usepackage{multirow}
\usepackage{verbatim}

\usepackage{youngtab}
\usepackage{young}

\def\bk{{\bold{k}}}

\setcitestyle{super}

\newcommand*{\citen}[1]{%
  \begingroup
    \romannumeral-`\x % remove space at the beginning of \setcitestyle
    \setcitestyle{numbers}%
    \cite{#1}%
  \endgroup   
}

\def\bk{{\bold{k}}}

\def\m1{{^{-1}}}

\newcommand{\clb}{\color{blue}}
\newcommand{\clr}{\color{red}}

\begin{document}

%%%%%%%%%%%%%%%%%%%%%%%%%%%%%%%%%%%%%%%%%
%%%%%%%%%%%%%%%%%%%%%%%%%%%%%%%%%%%%%%%%%
%%%%%%%%%%%%%%%%%%%%%%%%%%%%%%%%%%%%%%%%%

\title{Tailoring $T_c$ by symmetry principles: The concept of Superconducting Fitness}

\author{Aline Ramires}
\affiliation{Institute for Theoretical Studies, ETH Zurich, CH-8092, Zurich, Switzerland.}

\author{Daniel F. Agterberg}
\affiliation{Department of Physics, University of Wisconsin, Milwaukee, WI 53201, USA}

\author{Manfred Sigrist}
\affiliation{Institute for Theoretical Physics, ETH Zurich, CH-8093, Zurich, Switzerland}

\begin{abstract}

We propose a generalization of the concept of \emph{superconducting fitness}, which allows us to make statements analogous to \emph{Anderson's theorems} concerning the stability of different superconducting states. This concept can be applied to complex materials with several orbital, layer, sublattice or valley degrees of freedom. The superconducting fitness parameters $F_A(\bk)$ and $F_C(\bk)$ give a direct measure of the robustness of the weak coupling instability and of the presence of detrimental terms in the Hamiltonian, respectively. These two parameters can be employed as a guide to engineer normal state Hamiltonians in order to favour or suppress superconducting order parameters with different symmetries and topological properties. To illustrate the applicability and power of this concept we study three cases: the non-centrosymmetric heavy fermion CePt$_3$Si, the hole doped iron pnictide KFe$_2$As$_2$ and the doped topological insulator Cu$_x$Bi$_2$Se$_3$.

\end{abstract}

\date{\today}

\maketitle

%%%%%%%%%%%%%%%%%%%%%%%%%%%%%%%%%%%%%%%%%
%%%%%%%%%%%%%%%%%%%%%%%%%%%%%%%%%%%%%%%%%
%%%%%%%%%%%%%%%%%%%%%%%%%%%%%%%%%%%%%%%%%

%Intro: Why do we need something as the concept of SC Fitness?

Simple superconductors are usually well described by a single band of doubly degenerate electrons with only the spin degree of freedom (DOF). Their phenomenology can be addressed by strong symmetry-based arguments, such as Anderson's Theorems, which predict the vulnerability of different superconducting states through the presence of key symmetry breaking fields \cite{And1, And2} . The most interesting superconductors available today are in fact complex materials, which have extra DOFs, such as orbital \cite{Ram, Ram17, Fis, gao10,nic12,nou16-1,ong16,nic17,agt17,yon17, nom16,nom16-2, sch17,Agt}, layer \cite{Gor,bzd17}, sublattice or valley \cite{Wat,yan17,Nak,Xi,Man,Lu}. Many of them host unconventional superconducting states with phenomenology which seems to go beyond the intuition developed for simple systems. In particular, unexpected response under key symmetry breaking fields, such as unusual upper critical field anisotropy has been observed \cite{Ram, Ram17,Lu}. Effective models which do not carefully take into account the underlying symmetries and properties of the extra electronic DOF usually fail to describe their behaviour. This requires us to start with a microscopic description in the basis in which the all symmetry properties are explicit (here we refer to this basis as the local orbital basis, but it can be associated with other microscopic DOF). On the other hand, superconductivity is well understood as a weak-coupling Fermi surface instability, which can be arbitrarily complex and is described in the band basis. The robustness of the instability is ultimately guaranteed by the presence of the desired states to be paired at the Fermi energy with opposite momenta, which can be quantified by the \emph{superconducting fitness}\cite{Ram,Fis}. This concept captures the complexity of the electronic states and encodes all the symmetry aspects in a concise form, allowing one to make statements analogous to Anderson's Theorems for complex superconducting materials.

%%%%%%%%%%%%%%%%%%%%%%%%%%%%%%%%%%%%%%%%%
%This work

This paper generalizes the concept of superconducting fitness previously introduced by some of the authors\cite{Ram,Fis}, now within a non-perturbative scheme, extending the range of applicability. The original concept of superconducting fitness (characterized by the function $F_C(\bk)$ defined below), assumes there is a weak-coupling instability and provides a way to check how different kinds of perturbations would destabilize a given superconducting order parameter. From this point, valid questions are: How to guarantee the presence of a weak coupling instability to start with? Can we assure that there is \emph{some} component of the order parameter in the orbital basis which will lead to an intra-band order parameter in the band basis? The answer to these questions is yes, and here we introduce a second superconducting fitness function, $F_A(\bk)$, which allows us to assess the presence of a weak coupling instability. The generalization of the concept of superconducting fitness also allows us to better understand the role of each term in the normal state Hamiltonian, and how it supports or suppresses different kinds of superconducting order parameters. With this concept in hand, we have a very portable and inexpensive tool to, in principle, engineer the normal electronic state in order to support the emergence of specific superconducting order parameters, which can be exotic or topological. In this paper we apply this generalized concept to the non-centrosymmetric material CePt$_3$Si, the hole doped iron-pnictide K$_x$Fe$_2$As$_2$, and  the doped topological insulator Cu$_x$Bi$_2$Se$_3$.

%%%%%%%%%%%%%%%%%%%%%%%%%%%%%%%%%%%%%%%%%
%Two orbital problem

We focus here on a generic two-orbital model, a minimal model for the discussion of multi-band superconductivity. The Hamiltonian can be written in terms of an `orbital $\otimes$ spin' basis, $\Psi_{OS}^\dagger=(c_{1\uparrow}^\dagger,c_{1\downarrow}^\dagger,c_{2\uparrow}^\dagger,c_{2\downarrow}^\dagger )$, and parametrized as:
\begin{eqnarray}\label{H0h}
H_0(\bk)= \sum_{a,b} h_{ab}(\bk) \tau_a\otimes\sigma_b,
\end{eqnarray}
where the Pauli matrices $\tau_a$ and $\sigma_b$ correspond to the extra (orbital) and spin DOF, respectively. $h_{ab}(\bk)$ are real functions of momenta $\bk$ due to the hermiticity of the Hamiltonian. In presence of parity ($P: \bk\rightarrow -\bk$, for orbitals of same parity) and time-reversal symmetry (TRS) defined as $\Theta = K \tau_0\otimes (i\sigma_2)$, the only pairs allowed for $(a,b)$ are $(0,0),(1,0),(3,0),(2,1),(2,2),(2,3)$. For orbitals with opposite parity ($P: \tau_3 \otimes \sigma_0$ and $\bk\rightarrow -\bk$), the allowed pairs in presence of TRS are $(0,0), (2,0),(3,0),(1,1),(1,2),(1,3)$. For the case of a sublattice DOF, such as in graphene, parity also takes one sublattice into another ($P: \tau_1\otimes\sigma_0$ and $\bk\rightarrow -\bk$), and as a consequence the allowed pairs are $(0,0), (1,0),(2,0),(3,1),(3,2),(3,3)$. In all cases, the set of matrices $\tau_a\otimes\sigma_b$ is completely anti-commuting, an important property to simplify the equations below.

In order to introduce a more concise notation, we substitute the $\tau_a\otimes\sigma_b$ matrices by $T_i$-operators labelled by $i=1,...,15$, which can be identified with the generators of SU(4), traceless hermitian matrices, plus $T_0=\tau_0\otimes \sigma_0$. The generators follow the relation:
$T_i T_j = \delta_{ij} I_0 + \sum_c (i f_{ijk}+ d_{ijk})T_k,$
where $f_{ijk}= Tr[[T_i,T_j]T_k]$ and $d_{ijk} = Tr[\{T_i,T_j\} T_k]$ are completely anti-symmetric and symmetric structure constants, respectively. We distinguish the special subset of 5 completely anti-commuting generators by a different letter $O_i$, such that $T_{1...5}$ are identified with $O_{1...5}$. These follow $\{O_i, O_j\} = 2\delta_{ij} I_0$. The Hamiltonian can then be written as:
\begin{eqnarray}
H_0(\bk)&=& c_0(\bk) T_0 + \textbf{c}(\bk) \cdot \textbf{O},
\end{eqnarray}
where $c_0(\bk)$ is an even function of $\bk$ , $\textbf{O} = (O_1,...,O_{5})$ and $\textbf{c}(\bk)$ is a 5-dimensional vector. In presence of parity and TRS the problem is doubly degenerate, so we can parametrize the Green's functions (GF) in terms of two poles in a convenient way:
\begin{eqnarray}
G_0 (k)= G_a(k) \frac{( I_0 +  \hat{\textbf{c}}(\bk)\cdot \textbf{O})}{2} + G_b(k) \frac{( I_0 -  \hat{\textbf{c}}(\bk)\cdot\textbf{O})}{2},\nonumber\\
\end{eqnarray}
where $G_{a,b} (k) = G_{a,b}(\bk,i\omega_n)$ and $ \hat{\textbf{c}}(\bk) = \textbf{c}(\bk)/|\textbf{c}(\bk)|$. From $(i\omega_n T_0 - H_0)G_0(k) = T_0$ we find $G_{a,b}(k) = (i\omega_n - \epsilon_{a,b}(\bk))^{-1}$, where $\epsilon_{a,b}(\bk) = c_0(\bk) \pm |\textbf{c} (\bk)|$.

We can also write the order parameter in terms of the generators of SU(4) as a gap matrix $\Delta(\bk) = \bold{d}(\bk)\cdot \textbf{T} (i \Sigma_2)$, where $\bold{d}(\bk)$ is a 16-dimensional vector, $\textbf{T} = (T_0,T_1,...,T_{15})$ and $i\Sigma_2 = \tau_0 \otimes (i\sigma_2)$. Given the fact that the `two orbital $\otimes$ spin'  problem can also be thought in terms of a $j=3/2$ problem \cite{bry16,kim16,sav17}, we can analogously  classify the order parameter as singlet, triplet, quintuplet{\clr ,} and septuplet. The singlet is described by the identity matrix $T_0$, while the states in the quintuplet by the set of 5 (non-symplectic) generators introduced above in the context of orbitals with same parity, satisfying $\Sigma_2 O_i \Sigma_2 = O_i^T$. The states in the triplet and septuplet are described by the remaining 10 (symplectic) generators, following $\Sigma_2 T_i \Sigma_2 = -T_i^T$.  

We can then insert the parametrized GF and gap matrix in the linearized gap equation 
\begin{eqnarray}
1 = -T v \sum_{\bk , n } Tr[ \hat{\Delta}^\dagger(\bk)G_{0}(k) \hat{\Delta}(\bk) G^T_{0}(-k) ],
\end{eqnarray}
where $v$ is the magnitude of the attractive interaction in the symmetry channel of the respective $\Delta(\bk) = d \hat{\Delta}(\bk)$, where $\hat{\Delta}(\bk)$ is the normalized gap matrix satisfying $\langle | \hat{\Delta}(\bk)|^2\rangle_{FS} =1$.  After some manipulation we find the suggestive form:
\begin{eqnarray}\label{GapF}
1 &=& -\frac{T v}{8} \sum_{\bk , n} \Big[  (G_a\bar{G}_a + G_b\bar{G}_b)Tr[|F_A(\bk)|^2]\\ \nonumber &+& (G_a\bar{G}_b + G_b\bar{G}_a)Tr[|F_C(\bk)|^2]\Big],
\end{eqnarray}
where $G_{a,b} = G_{a,b}(k)$ and $\bar{G}_{a,b} = G_{a,b} (-k)$ and here we introduce the superconducting fitness parameters:
\begin{eqnarray}\label{DefF}
F_A(\bk) (i\sigma_2) =\tilde{ H}_0(\bk) \hat{\Delta}(\bk) + \hat{\Delta}(\bk) \tilde{H}_0^*(-\bk),\\ \nonumber
F_C(\bk) (i\sigma_2) =\tilde{ H}_0(\bk) \hat{\Delta}(\bk) - \hat{\Delta}(\bk) \tilde{H}_0^*(-\bk),\end{eqnarray}
where $\tilde{ H}_0(\bk)  = (H_0(\bk) - c_0(\bk)T_0 )/|\textbf{c}(\bk)|$. From the definition above it becomes clear that $Tr[|F_A(\bk)|^2]$ quantifies the presence of \emph{intra-band pairing}, therefore guaranteeing the robustness of the SC instability in the weak-coupling limit; while $Tr[|F_C(\bk)|^2]$ quantifies the presence of \emph{inter-band pairing}, usually associated with detrimental effects to the respective superconducting state \cite{Ram}.

The first line in Eq.~\ref{GapF} can be treated as the usual BCS equation, leading to the familiar form, now carrying the superconducting fitness parameter $F_A(\bk)$:
\begin{eqnarray}
&& -2v \alpha ln \left(\frac{4 e^\gamma}{\pi}\frac{\omega_C}{2 k_B T}\right),
\end{eqnarray}
where
\begin{eqnarray}
16 \alpha = \left[ N_a(0) \langle Tr[|F_A(\bk)|^2]\rangle_{FS_a} + (a\rightarrow b) \right],
\end{eqnarray}
$\omega_C$ is a characteristic cutoff energy, $N_{a,b}(0)$ are the density of states at the Fermi energy and $\langle ... \rangle_{FS_{a,b}}$ denotes the average over the Fermi surface for bands $a$ and $b$, respectively.

The second line, after the sum over the Matsubara frequencies, can be written as:
\begin{eqnarray}
 - \frac{ v}{8} \sum_{\bk}\frac{\text{tanh} \left(\frac{\beta \epsilon_a(\bk)}{2}\right)  }{(\epsilon_a(\bk)+\epsilon_b(\bk))} Tr[|F_C(\bk)|^2] + (a\rightarrow b).
\end{eqnarray}
At this point, for the single band scenario, one usually turns the integral over momenta into an integral over energy, introducing a DOS and integrating over a narrow range over the FS between $\pm \omega_C$. Note, though, that this integral is different because its denominator is written as a sum of two dispersions, and the integral is dominated by the region where $\epsilon_a+\epsilon_b = 0$, but for well separated bands, this condition is normally satisfied far away from the FS. We introduce the quantity $q(\bk)= \epsilon_a(\bk)-\epsilon_b(\bk)$ and assume well separated bands, such that $q(\bk)>> \omega_C$ in the range of energies close to the FS. Within this approximation, we find that the second line in  Eq.~\ref{GapF}  is equal to $2v\delta$, where
\begin{eqnarray}
\!\!\!\!\!\!\! 16 \delta = \frac{\omega_C^2}{2} \left[  N_a(0)  \left\langle \frac{Tr[|F_C(\bk)|^2]}{q(\bk_{Fa})^2} \right\rangle_{\!\!FS_a} \!\!\! \!\!+ (a\rightarrow b) \right].
\end{eqnarray}

Now the full gap equation can be concisely written as:
\begin{eqnarray}
1 &=&  -2v\alpha  ln \left(\frac{4 e^\gamma}{\pi}\frac{\omega_C}{2 k_B T}\right)  +2 v \delta,
\end{eqnarray}
and the critical temperature is explicitly obtained as:
\begin{eqnarray}
k_B T_C &=&  \frac{4 e^\gamma}{\pi}\frac{\omega_C}{2 } e^{-1/2|v| \alpha} e^{-\delta /\alpha},
\end{eqnarray}
which makes clear that the larger the $\alpha$, the larger the critical temperature, and if $\delta$ is finite $T_C$ is suppressed.

Note that this closed form for $T_C$ was now obtained non-perturbativelly, within the usual assumptions in the weak coupling limit of $T_C<<\omega_C$, under the requirement of $q(\bk_F)>> \omega_C$, which is realistic for many materials of interest.

%%%%%%%%%%%%%%%%%%%%%%%%%%%%%%%%%%%%%%%%%
%One-orbital problem

Within this non-perturbative formalism one can also find a closed form for $T_C$ for the single band scenario. Starting from $H_0 = \epsilon(\bk)\sigma_0 + \textbf{s}(\bk)\cdot \boldsymbol{\sigma}$, where $\textbf{s}(\bk)$ is a three-dimensional vector associated with an external symmetry breaking field, we find $\langle Tr[|F_A(\bk)|^2]+Tr[|F_C(\bk)|^2]\rangle _{FS}=8$ for any superconducting state. Performing a calculation in the same spirit as in \cite{Sig09,Fri}, the gap equation leads to:
\begin{eqnarray}
1&=& - 2v  N(0) ln \left(\frac{4 e^\gamma }{\pi} \frac{\omega_C}{2 k_B T_C}\right) \\ \nonumber &-& \frac{v}{8} N(0)  \langle 2 f(\rho_\bk)Tr[|F_C(\bk)|^2]\rangle_{FS},
\end{eqnarray}
where $\rho_\bk = |\bold{s}_\bk|/\pi T$ and $f(x) = 2 \text{Re} \sum_n[(2n+1+ i x)^{-1}-(2n+1)^{-1}]$. The maximum critical temperature is achieved when $|\bold{s}_\bk| \rightarrow 0$, such that:
\begin{eqnarray}
ln \left(\frac{T_C}{T_C^{Max}}\right) &=& \frac{1}{16}  \langle 2 f(\rho_\bk)Tr[|F_C(\bk)|^2]\rangle_{FS}.
\end{eqnarray}

Expanding in $\rho_\bk$ we find:
\begin{eqnarray}
T_C 
&\approx&T_C^{Max} \left(1 - \frac{7 \zeta(3)}{64 \pi^2}  \frac{\langle \rho_\bk^2 Tr[|F_C(\bk)|^2]\rangle_{FS}}{(T_C^{Max})^2} \right),
\end{eqnarray}
which is the same as Eq. 27 from our previous work \cite{Ram}, up to a change in the normalization of $F_C(\bk)$.

%%%%%%%%%%%%%%%%%%%%%%%%%%%%%%%%%%%%%%%%%
%Applications

Now we apply the analysis of the two superconducting fitness parameters to several superconducting materials in order to show the power of this concept and what kind of robust statements one can make concerning the stability of different superconducting states, or how to better engineer the normal state in order to support the most interesting or exotic order parameter.

%%%%%%%%%%%%%%%%%%%%%%%%%%%%%%%%%%%%%%%%%
\emph{CePt$_3$Si}: is a heavy fermion superconductor without inversion symmetry \cite{Bau}. It has a $T_C \approx 0.75K$ and an upper critical field $H_{c2}\approx 5T$, much larger than the Pauli limit, naively suggesting it is a triplet superconductor \cite{Bau}. Simply modelling this system as a single band superconductor\cite{Fri,Tan}, the absence of inversion symmetry leads to the presence of a Rashba-type spin-orbit coupling (SOC) of the form $H_{SOC} = \lambda \textbf{g}_\bk\cdot \boldsymbol{\sigma}$, where $\lambda$ is the strength of the SOC coupling and $\textbf{g}_\bk =  (-k_y,k_x,0)$. The normalized triplet order parameters in accordance with the $C_{4v}$ point group and the respective averages of $Tr[|F_C(\bk)|^2]$ and $Tr[|F_A(\bk)|^2]$ are summarized in Table I in the supplemental material (SM). Note that all the spin triplet states with $A_1$ symmetry are strongly suppressed as soon as SOC is turned on, given the large value of $Tr[|F_C(\bk)|^2]$ and the small (or zero \cite{sch17}) value of $Tr[|F_A(\bk)|^2]$. Interestingly enough, the last state within this family has a non-zero $Tr[|F_A(\bk)|^2]$, which always guarantees the presence of a weak coupling instability in presence of SOC. These aspects can be related to Fig. 1 in Tanaka et al.\cite{Tan}, where one can see all $A_1$ states are strongly suppressed in presence of SOC, with the last state displaying a small tail for large $\lambda$. The state with $A_2$ symmetry is completely robust in presence of SOC since $Tr[|F_C(\bk)|^2]=0$. For the states with $B_1$, $B_2$ and $E$ symmetries, $Tr[|F_A(\bk)|^2]=Tr[|F_C(\bk)|^2]$, which tells us that the effect of SOC is two-fold: it is partially detrimental (by a finite $Tr[|F_C(\bk)|^2]$), but also supports the presence of the weak coupling instability (by a finite $Tr[|F_A(\bk)|^2]$), therefore these states are suppressed with increasing $\lambda$, but much more slowly  than those in the $A_1$ family, as can be also observed in Tanaka et al.\cite{Tan}. In conclusion, it is easily inferred from the superconducting fitness parameters that all the order parameters are suppressed in presence of SOC, with the exception of $A_2$ channel. 

%%%%%%%%%%%%%%%%%%%%%%%%%%%%%%%%%%%%%%%%%
\emph{KFe$_2$As$_2$}: is a strongly hole doped iron-based superconductor with $T_C\approx 3 K$ \cite{Sat}. Recently it was proposed that s-wave superconductivity in the spin triplet $A_{2g}$ channel is realized once the Hund's coupling is larger than inter-orbital Hubbard repulsion in presence of SOC \cite{Vaf}. A minimal model with 2 orbitals in the basis $\Psi_\bk^\dagger = (d_{yz \uparrow }^\dagger, d_{yz \downarrow }^\dagger, d_{xz \uparrow }^\dagger,d_{xz \downarrow }^\dagger )$, can be written as $
H_0= a(\bk) \tau_0\otimes \sigma_0 + b(\bk) \tau_3 \otimes \sigma_0 + c(\bk) \tau_1 \otimes \sigma_0$, where $a(\bk)=\bk^2/2m - \mu$ and $b(\bk) = b k_x k_y$ characterize intra-orbital hopping, while $c(\bk)= c (k_x^2-k_y^2)$ characterizes inter-orbital hopping. For this model, even though the interaction in the $A_{2g}$ channel can be attractive, the pairing is purely inter-band (with order parameter $\Delta_{A_{2g}}(\bk) = d_{A_{2g}}(\bk) \tau_2\otimes \sigma_1$), therefore the superconducting instability is not robust since the pairing susceptibility is not logarithmically divergent \cite{Vaf}. From the perspective of the superconducting fitness, this is captured by the fact that $F_A(\bk)=0$ (the fitness parameters for all superconducting states are presented in Table II in the SM). Once SOC is turned on, $ H_{SOC} =  \lambda \tau_2\otimes \sigma_3$, the $A_{1g}$ and $A_{2g}$ channels mix, and the authors in \cite{Vaf} argue that a weak-coupling instability develops because now the component with $A_{1g}$ symmetry (with order parameter $\Delta_{A_{1g}}(\bk) = d_{A_{1g}}(\bk) \tau_0\otimes \sigma_2$) has intra-band character. From the  superconducting fitness it actually becomes evident that in fact pairing in the $A_{2g}$ channel itself develops a weak-coupling instability in presence of SOC since $F_A (\bk) \sim 8 \lambda \tau_0\otimes \sigma_2$ is finite, even before considering the admixture with a $A_{1g}$ component. As a consequence, the prefactor in the term carrying the Cooper logarithm is proportional to $\lambda^2$, in accordance with previous results \cite{Vaf}. So the ultimate reason why there is a weak-coupling instability in this scenario is SOC leading to a mixture of orbitals such that intra-band pairing in the $A_{2g}$ channel is now possible, not by the admixture of a $A_{1g}$ component to the order parameter. This is corroborated by the fact that this conclusion is unaffected by the presence of a strong repulsive U in the $A_{1g}$ channel \cite{Vaf}. We would like to note here that SOC has a different role than in the previous example: now SOC is important for the stabilization of a weak coupling instability.

\begin{figure}[t]
\begin{center}
\includegraphics[width=0.8\linewidth, keepaspectratio]{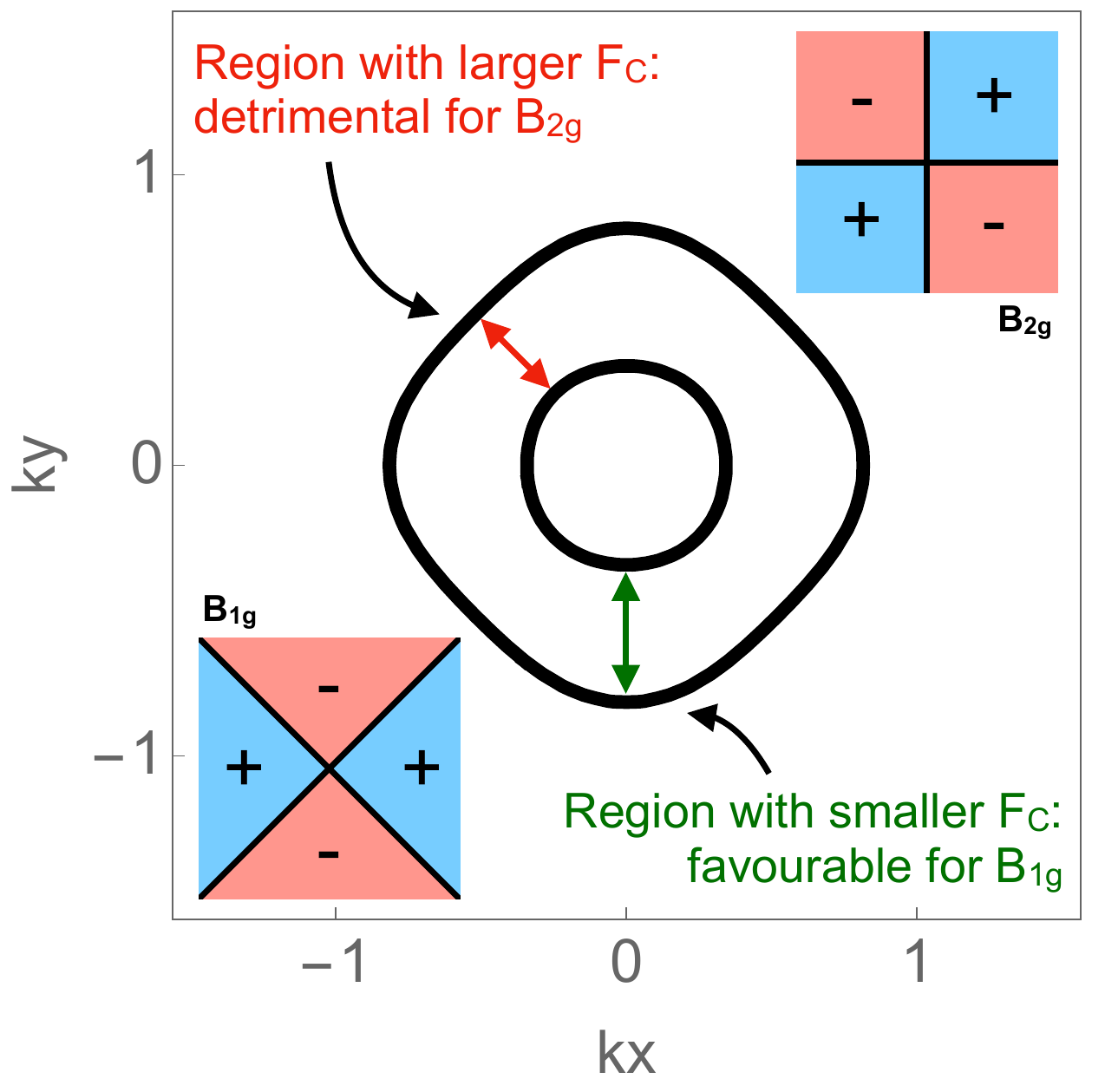}
\caption{Fermi surfaces from the effective model of KFe$_2$As$_2$. The insets display the form factors with $B_{1g}$ (left bottom) and $B_{2g}$ (top right) symmetries.}
\label{FeAs}
\end{center}
\end{figure}

The detailed analysis of the superconducting fitness parameters allows us to infer that SOC is generally detrimental to superconductors with d-wave character, while it supports states with s-wave symmetry. We can also discuss the competition between the two order parameters with d-wave symmetry. The SC instability in the $B_{1g} \sim k_x^2- k_y^2$  and $B_{2g} \sim k_x k_y$ channels is guaranteed by the presence of the $b(\bk)$  and $c(\bk)$ term, respectively. Note that the product of the form factors determining $Tr[|F_A(\bk)|^2]$ is the same for both channels. Ultimately, assuming the same pairing interaction in both channels, the magnitude of the constants $b$ and $c$ in the Hamiltonian is what determines which channel is favoured: in this case enhancing inter-orbital hopping favours the $B_{2g}$ channel. The parameter $Tr[|F_C(\bk)|^2]$ is also useful in determining the presence of detrimental effects: since it is proportional to $1/q(\bk_F)$, it is minimized when the Fermi surfaces are far apart in the Brillouin zone. As a consequence, the position of the nodes would tend to be in the direction where the two Fermi surfaces are the closest, in this case favouring the $B_{1g}$ channel, as schematically shown in Fig.~\ref{FeAs}. These arguments suggest interesting directions for engineering of superconducting states based in the application of strain.

%%%%%%%%%%%%%%%%%%%%%%%%%%%%%%%%%%%%%%%%%
\emph{C\MakeLowercase{u}$_x$B\MakeLowercase{i}$_2$S\MakeLowercase{e}$_3$}: is a candidate topological superconductor \cite{Fu}.  Zhang et al. \cite{Zha} proposed a two-orbital effective model based on two effective $p_z$-like orbitals which are a superposition of $p_z$ orbitals from Bi and Se. They label the orbitals as $P_{1z+}$ and $P_{2z-}$, with the $\pm$ sign indicating the parity of the effective orbital. In the basis $\Phi^\dagger  = (c_{1\uparrow}^\dagger, c_{1\downarrow}^\dagger, c_{2\uparrow}^\dagger, c_{2\downarrow}^\dagger)$, the Hamiltonian can be parametrized as in Eq.~\ref{H0h}, with non-zero $h_{ab}(\bk)$ only for $(a,b) = (0,0), (2,0),(3,0),(1,1),(1,2),(1,3)$ in presence of TRS and parity.  The explicit general form and character of these terms are summarized in Table III in the SM. The order parameter can also be parametrized as $\Delta = \sum_{a,b} d_{ab} (\bk) \tau_a \otimes \sigma_b$, and for the $D_{3d}$ point group symmetry there are four irreducible representations, displayed in Table IV in the SM. For simplicity we focus on $\bk$-independent order parameters. One can now evaluate the  superconducting fitness parameters for each term in the Hamiltonian separately. Table V in the SM summarizes which fitness parameters are zero or finite for each of the possible order parameters. Since all the terms in the Hamiltonian, when evaluated at the Fermi surface, have approximately the same magnitude (see Table III), with the exception of the $(1,3)$ term corresponding to trigonal warping, we can assign a score to each term as follows: a $-1$ score is given to each element in the Hamiltonian which contributes to a non-zero $F_C(\bk)$, while a $+1$ score is assigned to each element in the Hamiltonian contributing to a finite $F_A(\bk)$. These results are summarized in Table IV, and allows us to make the following observations: i) The dominant term in the Hamiltonian $(3,0)$ supports even and suppresses odd order parameters; ii) The terms $(1,1)$ and $(1,2)$, associated with SOC, support only the order parameters with $A_{1u}$ and $A_{2u}$ symmetry amongst the odd order parameters and the first order parameter with $A_{1g}$ symmetry; iii) The $(2,0)$ term supports order parameters in the $A_{1g}$ (first), $A_{1u}$ and $E_{u}$, while it supresses order parameters in the $A_{1g}$ (second) and $A_{2u}$ representations; iv) The trigonal warping term $(1,3)$ favours order parameters in the $A_{1g}$ (first), $A_{2u}$ and $E_{u}$ channels, while it suppresses order parameters in the $A_{1g}$ (second) and $A_{1u}$ channels. Fu and Berg \cite{Fu} proposed a criterium for the realization of topological SC which states: a time-reversal invariant, centro-symmetric superconductor is topological if it has odd-parity pairing with a full superconducting gap and if the Fermi surface encloses an odd number of time-reversal invariant momenta in the Brillouin zone. From the perspective of the superconducting fitness we can understand why the authors find that the most stable state with odd parity is in the $A_{1u}$ symmetry channel, since this is the only odd order parameter with only one sizeable detrimental term. We can now go further and propose how this material should be engineered such that an odd superconducting state is favoured. One direction is to make the $(3,0)$ term smaller. This term is associated with the difference in intra-orbital hopping for the two effective orbitals, and is dominated by in-plane hopping (see parameters in the SM and Ref.~\citen{Liu}). The study of the dependence of the magnitude of this parameter under different kinds of strain from DFT calculations would be insightful. Other terms can also be engineered in order to favour a specific odd parity order parameter, such as enhancement of the SOC, associated with $(1,1)$ and $(2,1)$ terms for the development of an order parameter with $A_{1u}$ or $A_{2u}$ symmetry. Interestingly enough, nematic superconductivity \cite{yon17,Fu14}, emerging from the two-dimensional $E_u$ representation is favoured by the enhancement of trigonal warping. 

In conclusion, we have introduced the concept of  superconducting fitness and the parameters $F_A (\bk)$ and $F_C (\bk)$ which quantify the robustness of the superconducting instability and the presence of detrimental effects, respectively. These two parameters provide a very handy and useful computational tool to evaluate the effects of each symmetry allowed term in the Hamiltonian and how it supports or suppresses different superconducting order parameters. Ultimately, this scheme can be used as a first guide to engineer complex materials in order to support the desired exotic superconducting states.

We thank Philip Brydon,  Mark Fischer, Dirk Manske, Mathias Scheurer, J\"org Schmalian, Carsten Timm,  and Jose Luis Lado Villanueva  for useful discussions. This work was partially supported by Dr. Max R\"{o}ssler, the Walter Haefner Foundation and the ETH Zurich Foundation (A.R.) and by the Swiss National Science Foundation (M.S.). D.F.A. acknowledges the hospitality of the Pauli Center of the ETH Zurich.

%%%%%%%%%%%%%%%%%%%%%%%%%%%%%%%%%%%%%%%%%
%%%%%%%%%%%%%%%%%%%%%%%%%%%%%%%%%%%%%%%%%
%%%%%%%%%%%%%%%%%%%%%%%%%%%%%%%%%%%%%%%%%
%Supplementary Information

\newpage

\begin{widetext}

\begin{table*}[p]
\label{TabCePt3Si}
\begin{center}
    \begin{tabular}{| c | c | c | c | c | c |}
    \hline
    Basis & Order Parameter ($\textbf{d}(\bk)$) & $Tr[|F_C(\bk)|^2]$ &  $Tr[|F_A(\bk)|^2]$ & Suppression of $T_c$ & Relation to SOC ($\textbf{g}_\bk$) \\ \hline
    \multirow{3}{*}{$A_1$} & $\sqrt{\frac{3}{2}}\frac{1}{k_F} (k_x,k_y,0)$ & $8 $ & 0  & Strong & $\bot$   \\ \cline{2-6}
    				       & $ \frac{1}{k_F} (k_x,k_y,k_z)$  & $8 $ &  0  & Strong &$\bot$\\ \cline{2-6}
  				       & $\sqrt{\frac{7}{2}}\frac{1}{k_F^3} (k_x^3,k_y^3,0)$ & $ 7.6 $ & $ 0.4 $ & Strong + Tail & other  \\ \hline
    $A_2$ & $\sqrt{\frac{3}{2}}\frac{1}{k_F} (-k_y,k_x,0)$  & 0 &  $8$  & No & $\parallel$ \\ \hline
    $B_1$ & $\sqrt{\frac{3}{2}}\frac{1}{k_F}  (k_x,-k_y,0)$ & $4 $  & $4 $ &Slight  & other \\ \hline
    $B_2$ & $\sqrt{\frac{3}{2}}\frac{1}{k_F}  (k_y,k_x,0)$& $4 $    & $4 $   &Slight & other\\ \hline
    \multirow{2}{*}{$E$}  & $\frac{\sqrt{3}}{k_F} (k_z,0,0)$ & $4$  & $4$ &Slight  & other\\ \cline{2-6}
                                 & $\frac{\sqrt{3}}{k_F} (0,k_z,0)$ & $4$ & $4$ &Slight  & other\\ \hline
    \end{tabular}
        \caption{Table summarizing the effects of SOC for the example of CePt$_3$Si. The table enumerates the allowed triplet superconducting states with $C_{4v}$ point group symmetry parametrized as $\hat{\Delta}(\bk) = \textbf{d}(\bk)\cdot\boldsymbol{\sigma}(i\sigma_2)$, the respective  superconducting fitness parameters, the suppression $T_C$ and the relation between the $\textbf{d}_\bk$ and $\textbf{g}_\bk$ vectors in each case.}
\end{center}
\end{table*}

\begin{table*}[p]
\label{TabFeAs}
\begin{center}
    \begin{tabular}{| c | c | c | c | c | c | c |}
    \hline
    Basis & Spin & Orbital & Parity & Matrix Form & $Tr[|F_C(\bk)|^2]$* &  $Tr[|F_A(\bk)|^2]$* \\ \hline
    $A_{1g}$ & Singlet & Intra & Even (s) &$\tau_0\otimes(i\sigma_2)$  &  $0$  &  $|d_{A_{1g}}(\bk)|^2(| b(\bk)|^2 +| c(\bk)|^2 +  \lambda^2)$  \\ \hline
    $B_{1g}$ & Singlet & Intra & Even (d)& $\tau_3\otimes(i\sigma_2)$ &  $|d_{B_{1g}}(\bk)|^2(\lambda^2+|c({\bf k})|^2)$  &  $|d_{B_{1g}}(\bk)|^2| b(\bk)|^2$\\ \hline
    $B_{2g}$ & Singlet & Inter & Even (d)& $\tau_1\otimes(i\sigma_2)$&  $ |d_{B_{2g}}(\bk)|^2(\lambda^2+|b({\bf k})|^2)$    & $|d_{B_{2g}}(\bk)|^2| c(\bk) |^2$ \\ \hline
    $A_{2g}$ & Triplet & Inter & Even (s)& $\tau_2\otimes\sigma_1$  & $|d_{A_{2g}}(\bk)|^2 (| c(\bk)|^2+|b({\bf k})|^2$) &  $  |d_{A_{2g}}(\bk)|^2 \lambda^2$ \\ \hline
    \end{tabular}
        \caption{Table summarizing the superconducting fitness parameters of different superconducting states for the example of KFe$_2$As$_2$. * Values should be normalized by $(| b(\bk)|^2+ | c(\bk)|^2 + \lambda^2)/64$, where $b(\bk) = b k_x k_y$ and $c(\bk)= c (k_x^2-k_y^2)$, characterizing intra- and inter-orbital hopping, repectively. The momentum dependency of the order parameters are the following: $d_{A_{1g}} \sim \text{cte}$, $d_{A_{2g}} \sim \text{cte}$, $d_{B_{1g}} \sim (k_x^2-k_y^2)$ and $d_{B_{2g}} \sim k_x k_y$. In order to reproduce the Fermi surface in Fig. 1 in the main text, the following parameters were used: $m=1$, $b=0.63$, $c=0.35$, $\mu=0.25$ \cite{Vaf}.}
\end{center}
\end{table*}

\begin{table*}[p]
\begin{center}
    \begin{tabular}{| c | c | c | c |}
    \hline
    $(a,b)$ & General form & Magnitude (eV) & Character   \\ \hline
    $(0,0)$ & $a_0 + a_p (k_x^2 + k_y^2)+ a_z k_z^2$ & $\approx 0.3$ & Intra-orbital hopping \\ \hline
    $(2,0)$ & $b_1 k_z+ b_3  k_z^3 + b_T k_y(3 k_x^2-k_y^2)$ &  $\approx 0.2$ &  Inter-orbital hopping\\ \hline
    $(3,0)$ & $c_0 + c_p (k_x^2 + k_y^2)+ c_z k_z^2$ & $ \approx 0.45 $ &  Intra-orbital hopping \\ \hline
    $(1,1)$ & $d_1 k_x + d_2k_x (k_x^2 + k_y^2)$ & $\approx 0.3$ & SOC \\ \hline
    $(1,2)$ & $ -d_1 k_y - d_2 k_y(k_x^2 + k_y^2)$&$  \approx 0.3$ & SOC \\ \hline
    $(1,3)$ & $e_T k_x(k_x^2-3k_y^2)$ & $\approx 0.05$ & Trigonal warping \\ \hline
    \end{tabular}
    \caption{Symmetry allowed terms for the Hamiltonian $H_0(\bk)= \sum_{a,b} h_{ab}(\bk) \tau_a\otimes\sigma_b,$  of Cu$_x$Bi$_2$Se$_3$ with their momentum dependence up to $O(\bk^2)$, the respective order of magnitude at the Fermi surface and the character of the term. For the magnitudes above we used the maximum value of the parameters cited in Liu et al.\cite{Liu} at the Fermi surface with $k_F \approx 0.1 \AA$ (this can be obtained from the ARPES experiment \cite{Wra} or by the value of the electron density upon doping\cite{Sas} $
\sim 10^{20} cm^{-3}$). Note that all but the last trigonal warping term are of the same order of magnitude.}
\end{center}
\end{table*}

\begin{table*}[p]
\begin{center}
    \begin{tabular}{| c | c | c | c | c | c | c  | }
    \hline
    Irrep &  Spin &  Orbital   & Parity   & Matrix Form &  $\langle |F_C(\bk)|^2\rangle_{FS}$ & $\langle |F_A(\bk)|^2\rangle_{FS}$ \\ \hline
     \multirow{2}{*}{$A_{1g}$}&
     \multirow{2}{*}{Singlet}&
          \multirow{2}{*}{Intra}&
               \multirow{2}{*}{Even}    & $\tau_0\otimes(i\sigma_2)$& 0 & +4   \\ \cline{5-7}
     					&&&& $\tau_3\otimes(i\sigma_2)$    &-3 & +1 \\ \hline
		{$A_{1u}$}  & Triplet  & Inter & Odd  & $(i \tau_2) \otimes\sigma_1$ &-1 & +3   \\ \hline
		{$A_{2u}$}  & Singlet  & Inter & Odd  & $\tau_1\otimes(i\sigma_2)$ &-2 & +2 \\ \hline
		 \multirow{2}{*}{$E_{u}$}&
     \multirow{2}{*}{Triplet}&
          \multirow{2}{*}{Inter}&
               \multirow{2}{*}{Odd}  
				&$(i\tau_2)\otimes\sigma_0$  &-2 & +2  \\ \cline{5-7}
					&&&& $(i\tau_2)\otimes\sigma_3$  &-2 & +2   \\ \hline
    \end{tabular}
\end{center}
        \caption{Table summarizing the score for superconducting fitness parameters of different superconducting states for the example of Cu$_x$Bi$_2$Se$_3$. The previous to last column gives a $-1$ score to each element in the Hamiltonian which contributes to a non-zero $F_C(\bk)$, while the last column gives a $+1$ score to each element in the Hamiltonian contributing to a finite $F_A(\bk)$ (see table below). The table does not count the contribution from the trigonal warping term (1,3) since this is one order of magnitude smaller than the other terms.}
\end{table*}

\begin{table*}[p]
    \begin{tabular}{| c | c | c | }
    \hline
    \multicolumn{3}{| c | }{$A_{1g}, (c,d) = (0,2)$} \\ \hline
    $(a,b)$ & $F_A(\bk) $&   $F_C(\bk)  $  \\ \hline
{\clb $(2,0)$} & $\neq 0$ & $=0$\\ \hline
{\clb $(3,0)$} &  $\neq 0$& $=0$\\ \hline
{\clb $(1,1)$} & $\neq 0$ & $=0$\\ \hline
{\clb $(1,2)$} &  $\neq 0$ &$=0$\\ \hline
{\clb $(1,3)$} & $\neq 0$ &$=0$\\ \hline
    \end{tabular}
    \begin{tabular}{| c | c | c | }
    \hline
    \multicolumn{3}{| c | }{$A_{1g}, (c,d) = (3,2)$} \\ \hline
    $(a,b)$ & $F_A(\bk) $&   $F_C(\bk)  $  \\ \hline
{\clr $(2,0)$} &  $= 0$ & $\neq0$\\ \hline
{\clb $(3,0)$} &  $\neq 0$ & $=0$\\ \hline
{\clr    $(1,1)$} &  $=0$ & $\neq0$\\ \hline
{\clr    $(1,2)$}&   $=0$ & $\neq 0$\\ \hline
{\clr    $(1,3)$}& $ = 0$ & $\neq0$\\ \hline
    \end{tabular}
    \begin{tabular}{| c | c | c | }
    \hline
    \multicolumn{3}{| c | }{$A_{1u}, (c,d) = (2,1)$} \\ \hline
    $(a,b)$ & $F_A(\bk) $&   $F_C(\bk)  $  \\ \hline
{\clb $(2,0)$} &  $\neq 0$ &$=0$\\ \hline
{\clr $(3,0)$} &  $= 0$ & $\neq0$\\ \hline
{\clb    $(1,1)$} &  $\neq 0$ & $=0$\\ \hline
{\clb    $(1,2)$}&   $\neq 0$ & $= 0$\\ \hline
{\clr     $(1,3)$} & $ = 0$ & $\neq0$\\ \hline
    \end{tabular}\\
    \begin{tabular}{| c | c | c | }
    \hline
    \multicolumn{3}{| c | }{$A_{2u}, (c,d) = (1,2)$} \\ \hline
    $(a,b)$ & $F_A(\bk) $&   $F_C(\bk)  $  \\ \hline
{\clr $(2,0)$} & $= 0$ & $\neq0$\\ \hline
{\clr $(3,0)$}&  $= 0$ & $\neq0$\\ \hline
{\clb $(1,1)$}&  $\neq 0$ & $=0$\\ \hline
{\clb $(1,2)$} & $\neq 0$ & $=0$\\ \hline
{\clb $(1,3)$}& $ \neq 0$ & $=0$\\ \hline
    \end{tabular}
    \begin{tabular}{| c | c | c | }
    \hline
    \multicolumn{3}{| c | }{$E_{u}, (c,d) = (2,0)$} \\ \hline
    $(a,b)$ & $F_A(\bk) $&   $F_C(\bk)  $  \\ \hline
{\clb$(2,0)$} &  $\neq 0$& $=0$\\ \hline
{\clr $(3,0)$}&  $= 0$ & $\neq0$\\ \hline
{\clb    $(1,1)$} &  $\neq 0$ & $=0$\\ \hline
{\clr     $(1,2)$} &   $= 0$ & $\neq  0$\\ \hline
{\clb    $(1,3)$} & $ \neq 0$ & $=0$\\ \hline
    \end{tabular}
     \begin{tabular}{| c | c | c | }
    \hline
    \multicolumn{3}{| c | }{$E_{u}, (c,d) = (2,3)$} \\ \hline
    $(a,b)$ & $F_A(\bk) $&   $F_C(\bk)  $  \\ \hline
{\clb$(2,0)$} &  $\neq 0$& $=0$\\ \hline
{\clr $(3,0)$}&  $= 0$ & $\neq0$\\ \hline
{\clr     $(1,1)$} &  $= 0$ & $\neq 0$\\ \hline
{\clb    $(1,2)$} &   $\neq 0$ & $=  0$\\ \hline
{\clb    $(1,3)$} & $ \neq 0$ & $=0$\\ \hline
    \end{tabular}
       \caption{Table summarizing when the superconducting fitness parameters are zero or finite for all the symmetry allowed terms in the Hamiltonian of Cu$_x$Bi$_2$Se$_3$, for all possible superconducting order parameters. These results were used to generate the scores in Table IV above. In blue we highlight the terms $(a,b)$ in the Hamiltonian which guarantee the weak coupling instability without introducing any detrimental effect, while in red we highlight the terms which only lead to a suppression of the critical temperature for the respective order parameter.}
\end{table*}

\end{widetext}

%%%%%%%%%%%%%%%%%%%%%%%%%%%%%%%%%%%%%%%%%
%%%%%%%%%%%%%%%%%%%%%%%%%%%%%%%%%%%%%%%%%
%%%%%%%%%%%%%%%%%%%%%%%%%%%%%%%%%%%%%%%%%

\end{document}